\documentstyle[12pt]{article}

\catcode`\@=11
\long\def\@makefntext#1{
\protect\noindent \hbox to 3.2pt {\hskip-.9pt
$^{{\ninerm\@thefnmark}}$\hfil}#1\hfill}		

\def\@makefnmark{\hbox to 0pt{$^{\@thefnmark}$\hss}}  

\def\ps@myheadings{\let\@mkboth\@gobbletwo
\def\@oddhead{\hbox{}
\rightmark\hfil\ninerm\thepage}
\def\@oddfoot{}\def\@evenhead{\ninerm\thepage\hfil
\leftmark\hbox{}}\def\@evenfoot{}
\def\sectionmark##1{}\def\subsectionmark##1{}}

\renewcommand{\thefootnote}{\fnsymbol{footnote}}

\newcounter{sectionc}\newcounter{subsectionc}\newcounter{subsubsectionc}
\renewcommand{\section}[1] {\vspace*{0.6cm}\addtocounter{sectionc}{1}
\setcounter{subsectionc}{0}\setcounter{subsubsectionc}{0}\noindent
	{\normalsize\bf\thesectionc. #1}\par\vspace*{0.4cm}}
\renewcommand{\subsection}[1] {\vspace*{0.6cm}\addtocounter{subsectionc}{1}
	\setcounter{subsubsectionc}{0}\noindent
	{\normalsize\it\thesectionc.\thesubsectionc. #1}\par\vspace*{0.4cm}}
\renewcommand{\subsubsection}[1]
{\vspace*{0.6cm}\addtocounter{subsubsectionc}{1}
	\noindent {\normalsize\rm\thesectionc.\thesubsectionc.\thesubsubsectionc.
	#1}\par\vspace*{0.4cm}}

\newcounter{appendixc}
\newcounter{subappendixc}[appendixc]
\newcounter{subsubappendixc}[subappendixc]

\renewcommand{\appendix}[1] {\vspace*{0.6cm}
        \refstepcounter{appendixc}
        \setcounter{figure}{0}
        \setcounter{table}{0}
        \setcounter{equation}{0}
        \renewcommand{\thefigure}{\Alph{appendixc}.\arabic{figure}}
        \renewcommand{\thetable}{\Alph{appendixc}.\arabic{table}}
        \renewcommand{\theappendixc}{\Alph{appendixc}}
        \renewcommand{\theequation}{\Alph{appendixc}.\arabic{equation}}
        \noindent{\bf Appendix \theappendixc #1}\par\vspace*{0.4cm}}

\renewenvironment{thebibliography}[1]
	{\begin{list}{\arabic{enumi}.}
	{\usecounter{enumi}\setlength{\parsep}{0pt}
\setlength{\leftmargin 1.25cm}{\rightmargin 0pt}
	 \setlength{\itemsep}{0pt} \settowidth
	{\labelwidth}{#1.}\sloppy}}{\end{list}}

\newcounter{itemlistc}
\newcounter{romanlistc}
\newcounter{alphlistc}
\newcounter{arabiclistc}

\newcommand{\fcaption}[1]{
        \refstepcounter{figure}
        \setbox\@tempboxa = \hbox{\footnotesize Fig.~\thefigure. #1}
        \ifdim \wd\@tempboxa > 6in
           {\begin{center}
        \parbox{6in}{\footnotesize\baselineskip=12pt Fig.~\thefigure. #1}
            \end{center}}
        \else
             {\begin{center}
             {\footnotesize Fig.~\thefigure. #1}
              \end{center}}
        \fi}

\newcommand{\tcaption}[1]{
        \refstepcounter{table}
        \setbox\@tempboxa = \hbox{\footnotesize Table~\thetable. #1}
        \ifdim \wd\@tempboxa > 6in
           {\begin{center}
        \parbox{6in}{\footnotesize\baselineskip=12pt Table~\thetable. #1}
            \end{center}}
        \else
             {\begin{center}
             {\footnotesize Table~\thetable. #1}
              \end{center}}
        \fi}

\def\@citex[#1]#2{\if@filesw\immediate\write\@auxout
	{\string\citation{#2}}\fi
\def\@citea{}\@cite{\@for\@citeb:=#2\do
	{\@citea\def\@citea{,}\@ifundefined
	{b@\@citeb}{{\bf ?}\@warning
	{Citation `\@citeb' on page \thepage \space undefined}}
	{\csname b@\@citeb\endcsname}}}{#1}}

\newif\if@cghi
\def\cite{\@cghitrue\@ifnextchar [{\@tempswatrue
	\@citex}{\@tempswafalse\@citex[]}}
\def\citelow{\@cghifalse\@ifnextchar [{\@tempswatrue
	\@citex}{\@tempswafalse\@citex[]}}
\def\@cite#1#2{{$\null^{#1}$\if@tempswa\typeout
	{IJCGA warning: optional citation argument
	ignored: `#2'} \fi}}

 1
 1
 1

\font\ninerm=cmr9

\newcommand{\ep}{\epsilon}

\newcommand{\la}{\lambda}

\newcommand{\La}{\Lambda}

\newcommand{\be}{\begin{eqnarray}}
\newcommand{\ee}{\end{eqnarray}}

\newcommand{\ti}{\tilde}

\newcommand{\ran}{\rangle}

\newcommand{\rar}{\rightarrow}

\begin{document}

\vspace{-25mm}
\begin{flushright}G\"oteborg ITP 95-07\end{flushright}

\centerline{\normalsize\bf States in the BRST cohomology for
{\it G/H} WZNW models.
\footnote{Talk presented by H.R. at "STRINGS '95", Los Angeles, March 1995.}}
\baselineskip=16pt

\centerline{\footnotesize Stephen Hwang}
\centerline{\footnotesize and}
\centerline{\footnotesize Henric Rhedin}
\baselineskip=13pt
\centerline{\footnotesize\it Institute of theoretical physics,
Chalmers University of Technology and G\"oteborg University}
\baselineskip=12pt
\centerline{\footnotesize\it S-412 96 G\"oteborg, Sweden}
\centerline{\footnotesize E-mail: tfesh@fy.chalmers.se, hr@fy.chalmers.se}

\vspace*{0.6cm}
\normalsize\baselineskip=15pt
\setcounter{footnote}{0}
\renewcommand{\thefootnote}{\alph{footnote}}

WZNW models, especially gauged WZNW models, are important in the study of
conformal field theories. Karabali and Schnitzer\cite{KS} initiated
the study of the BRST cohomology of
a WZNW model gauged by an anomaly free vector sub-group
and results were given for abelian sub-groups.
This result was generalized to non-abelian sub-groups for a specific
set of representations \cite{HR1}. The subject of this talk is the
analysis of arbitrary representations \cite{HR2,Hw}.

The effective action of a $G/H$ gauged WZNW model decomposes into three
sectors; the original WZNW model based on the group $G$, an auxiliary
WZNW model based on the sub-group $H$, and a Fadeev-Popov ghost part.
We assume that the state space decomposes into a direct product of
modules $M_{\la}^{h}\otimes\widetilde{M}_{\ti{\la}}^{h}\otimes{\cal F}$
where $M_{\la}^{h}$ and $\widetilde{M}_{\ti{\la}}^{h}$ are Verma modules
over affine Lie algebras $\hat{h}_k$ and $\hat{h}_{\ti{k}}$ with levels
$k$ and $\ti{k}=-k-2c_h$. $c_h$ is the quadratic Casimir in the
adjoint representation of $h$, and $\la$ and $\ti{\la}$ are weights of
the two sectors. ${\cal F}$ is a ghost Fock module.

Our analysis is based on the result that for an irreducible Verma
module, the BRST cohomology is confined to ghost free states \cite{HR1},
while in the reducible case we have for negative ghost numbers
the equation $\hat{Q}|S\ran=|N\ran$ for states $|S\ran$ in the
cohomology \cite{HR2,Hw}. Here $|N\ran$ is a null-vector of the Verma module.
The Verma module at hand is reducible for generic weights due
to prescence of null-vectors.

In order to analyse a reducible state space we perturb the weights
of the Verma modules by
an infinitesimal piece $\ep$ \cite{Ja} which gives an irreducible
state space. This will induce a gradation
of the Verma modules $M_{\la}^h=M^{(0)}\supset M^{(1)}\supset ...$ where
$M^{(i)}$ is the submodule which includes all states which are divisible
by $\ep ^i$ in the perturbed module. We define the irreducible sub-module
$L^h_{\la}\equiv M^{(0)}/M^{(1)}$. The states in $M^{(1)}/M^{(2)}$ are denoted
first generation null-vectors.

Using the perturbed Verma modules the relative cohomology is found to be
trivial for non zero ghost number, and only singlets and quartets will
be present \cite{HR2,Hw}. Here we use the terminology of Kugo and Ojima
\cite{KO}, where singlets and singlet pairs represent non-trivial
states in the cohomology, while quartets are sets of four states
that will not belong to the cohomology.
Now, if singlet pairs are generated as the
perturbation is set to zero, then they must appear from quartets, and
one must therefore identify suitable quartets. For negative ghost numbers
one finds that $\hat{Q}|\bar{S}_{-q}\ran=|\bar{N}_{-q+1}\ran$ and for positive
ghost numbers $\hat{Q}|N_{q-1}\ran_{\ep}=\ep|S_q\ran_{\ep}$ \cite{HR2,Hw}.
Here $\hat{Q}$ is the relative BRST operator, $|\bar{S}_{-q}\ran$ and
{\it lim}$_{\ep\rar 0}|S_q\ran_{\ep}$ is the singlet pair, and
$|\bar{N}_{-q+1}\ran$, {\it lim}$_{\ep\rar 0}|N_{q-1}\ran_{\ep}$ are first
generation null vectors. Subscript $\ep$ indicates that those
states are perturbed as described above. Furthermore
$|\bar{N}_{-q+1}\ran$ and {\it lim}$_{\ep\rar 0}|N_{q-1}\ran_{\ep}$ are states
in the cohomology in the irreducible sub-module of $M^{(1)}$.
Starting at ghost number zero we may now iteratively solve the equations
$\hat{Q}|\bar{S}_{-q}\ran=|\bar{N}_{-q+1}\ran$ and
$\hat{Q}|N_{q-1}\ran_{\ep}=\ep|S_q\ran_{\ep}$
using explicit expressions for singular vectors \cite{MFF}.

For ghost number plus and minus one the null-states
$|\bar{N}_{0}\ran$ and {\it lim}$_{\ep\rar 0}|N_{0}\ran_{\ep}$ are
constructed from known states in the cohomology at ghost
number zero \cite{HR1,Hw} by
substituting highest weight primaries by highest weight null-vectors.
One now solves $\hat{Q}|\bar{S}_{-1}\ran=|\bar{N}_{0}\ran$ and
$\hat{Q}|N_{0}\ran_{\ep}=\ep|S_1\ran_{\ep}$ to obtain the singlet pair
$|\bar{S}_{-1}\ran$ and {\it lim}$_{\ep\rar 0}|S_1\ran_{\ep}$. States in the
cohomology at ghost number plus and minus one are now used to get
$|\bar{N}_{-1}\ran$ and {\it lim}$_{\ep\rar 0}|N_{1}\ran_{\ep}$ in the same
manner as above etc.

It is important to note that for positive ghost numbers it is sufficient
to use either the original or the auxiliary sector, while for negative
ghost numbers
both sectors are required. Also, exchanging the sectors for positive
ghost numbers yields cohomologically equivalent results \cite{Hw}.

This construction yields singlet
pairs in the irreducible sub-module $L_{\la}^h\otimes\ti{L}_{\ti{\la}}^h$
with ghost numbers
given by $p=\pm|l_{\la}-l_{\ti{\la}}|$ where $l_{\la}$ is the minimum number
of Weyl reflexions taking us to $\la$ starting from a dominant weight $\mu$.
$l_{\ti{\la}}$ is defined in an analogous manner.

This construction remains completely valid for topological models as well
as for integrable representations of the original sector.
For the generic $G/H$ model
the assumption that the Verma module $M^g_{\La}$
decomposes into Verma modules $M^h_{\la}$ does not
always hold and some states may not stay in the cohomology. In fact,
all singlet pairs for which the construction involves
null-vectors of the affine Lie
algebra $\hat{h}_k$ that are not null-vectors of the full affine Lie algebra
$\hat{g}_k$, will not be in the cohomology. Since the construction is
iterative this may in fact mean that infinitely many ghost numbers vanish.

\end{document}